\documentclass[sn-mathphys]{sn-jnl}% Math and Physical Sciences Reference Style
\jyear{2021}%
\usepackage{amsmath}
\usepackage{amsthm}
\usepackage{amsfonts}
\usepackage{listings}
\usepackage{tabto}
\usepackage{physics}
\usepackage{placeins}

\theoremstyle{thmstyleone}%
\newtheorem{theorem}{Theorem}%meant for continuous numbers
\theoremstyle{thmstyletwo}%

\theoremstyle{thmstylethree}%

\begin{document}

\title[Quantum TV]{Quantum median filter for Total Variation image denoising}


\author[1,2]{\fnm{Simone} \sur{De Santis}}\email{simone.desantis@studio.unibo.it}

\author[1]{\fnm{Damiana} \sur{Lazzaro}}\email{damiana.lazzaro@unibo.it}

\author[2]{\fnm{Riccardo} \sur{Mengoni}}\email{r.mengoni@cineca.it}

\author*[1]{\fnm{Serena} \sur{Morigi}}\email{serena.morigi@unibo.it}

\affil*[1]{\orgdiv{Department of Mathematics}, \orgname{University of Bologna}, \orgaddress{\street{P.zza Porta San Donato}, \city{Bologna}, \postcode{40126}, \country{Italy}}}

\affil[2]{\orgdiv{CINECA}, \orgname{Organization}, \orgaddress{\street{Via Magnanelli 6/3,
Casalecchio di Reno}, \city{Bologna}, \postcode{40033},  \country{Italy}}}


\abstract{In this new computing paradigm, named quantum computing, researchers from all over the world are taking their first steps in designing quantum circuits for image processing, through a difficult process of knowledge transfer. This effort is named Quantum Image Processing, an emerging research field pushed by powerful parallel computing capabilities of quantum computers. This work goes in this direction and proposes the challenging development of a powerful method of image denoising, such as the Total Variation (TV) model, in a quantum environment. The proposed Quantum TV is described and its sub-components are analysed. Despite the natural limitations of the current capabilities of quantum devices, the experimental results show a competitive denoising performance compared to the classical variational TV counterpart.}




\maketitle

\section{Introduction}\label{sec1}

Digital image processing is an extremely, computationally demanding, strategic research field.
This has led, over the years, to the development of many complex image processing algorithms on highly parallel,  specialized hardware platforms. With the rapid progress of 
parallel hardware, suitably high performance is now available  also for sophisticated imaging tasks. 
In the present study we focus on image denoising, 
an essential requirement for any other image processing, which refers to the recovery of a clean sharp image from a noisy observation.
%area that finds application and utility in many fields, and it has gained relevance by providing powerful tools for different needs. Over the years, researchers have been designing algorithms to enhance image quality. They've been used for many purposes (i.e. art restoration, gaming and machine learning) and some of them have reached high performances thanks to technical achievements in parallel computation.

In \cite{ROF92}  Rudin, Osher and Fatemi presented one of the main mathematical models and algorithms for image denoising, the so-called Total Variation denoising (also known as \emph{TV model}). Since then, TV model has become a quite popular technique, which usage allows to improve overall image quality when the images are affected by noise or corruption, while well preserving edges and details. Efficient solutions have been proposed over time for the numerical optimizations of the TV model \cite{CTY13,CP}, but they are not able to fully exploit parallel computational resources. In \cite{OsherTV}, the authors designed a new optimization algorithm which is simple and highly parallelizable, and relies on median value computations, thus reducing computational effort to a sorting problem.
Thanks to its low complexity, this algorithm is prone to be implemented on low-end devices or, more generally, in situations where a reduced amount of resources is available. That is the case of Quantum Image Processing (QIP), a novel and promising research field which goal is the development of image processing techniques for quantum computers, exploiting peculiar features from quantum world, like entanglement, superposition, interference\cite{review1}. Quantum computing is universally acknowledged for its ability to process data providing computational speedups compared to the classical paradigm. However, image processing is actually one of the most demanding applications in terms of resources for quantum computers.
As a consequence QIP is still in its early stage, and thus is facing several fundamental problems, such as
how to represent and store an image in quantum computers appropriately, and how to
efficiently implement image processing algorithms \cite{QM,face}. 

In the present days, quantum devices are subject to several issues (i.e. noise, absence of error correction, low amount of available qubits, etc.), which defines what researchers commonly refer to as \emph{Noise intermediate-scale quantum} era, or just NISQ \cite{NISQ}.  This is an intermediate development phase where, through the exhibition of quantum devices' limits, algorithms have been designed to be as optimized and simple as possible \cite{NISQ_1}. In this context, QIP is a complex subject to deal with because several difficulties may arise in the attempt of implementing  the classical image processing algorithms on quantum devices (some of them will be discussed later in this work). Many proposals have been submitted during the past few years \cite{review2}, in the attempt to provide QIP algorithms that fulfill NISQ requirements. QIP algorithms are usually tested using quantum simulators or similar execution environments.

Here along this line, the presented work aims to solve  the TV model in a quantum environment. The result is a Quantum TV filter, which integrates a Quantum Median Filter proposal  by Li et al. \cite{QM_orig,QM}. This work focuses not only on the development of a QIP technique for image denoising which mimics its variational TV counterpart, 
but also on presenting an useful review of detected problems in the newly developed QIP field, and offering possible solutions and ideas for future developments and optimizations.

The paper is organized as follows. In Section \ref{sec:sec1} we introduce basic theory about Total Variation denoising technique and a median formula for efficiently solving an anisotropic TV problem. 
In Section \ref{sec:sec3} we discuss Quantum Image Processing concepts and related issues.
In Section \ref{sec:sec4} we explain in detail how Quantum TV Filter works, highlighting its quantum components, and analysing its circuit complexity in Section \ref{sec:sec5}.
In Section \ref{sec:sec6} we focus on the analysis of experimental results, comparing both quantum and classical implementations of the TV algorithm. Section \ref{sec:sec7} reports conclusions and future works.
In the Appendix we briefly report details on a few quantum modules
used in the design of the proposed Quantum TV.

For basic notations and fundamental knowledge about quantum image processing computing, we refer the reader to \cite{review1}.

\section{Total Variation image denoising}
\label{sec:sec1}

%In this chapter we focus on an image denoising technique, TV regularization; later we introduce a formula, which simplifies denoising computation by using an anisotropic model.

%For the content of this introduction to Total Variation denoising, we have referred to the paper \cite{Lanza} by Lanza et al.\\

The goal of denoising is to obtain an image $u^*$ not only with small variations in intensity between pixels but also close to the observation $f$. At this aim, the class of \emph{variational} methods for image restoration relies on determining
restored images $u^*\in\mathbb{R}^N$, given a noisy image $f\in\mathbb{R}^N$, as the minimizers of suitable cost functionals $J: \mathbb{R}^N \to \mathbb{R}$ such that, 
typically, restoration is casted as an optimization problem of the form:
\begin{equation}
u^* \:\;{\leftarrow}\;\: \arg \min_{u \in \mathbb{R}^N}
\left\{ \,
J(u) \;{:=}\; R(u) \;{+}\; \lambda \, F(u;f)
\, \right\}
\, ,
\label{eq:GVM}
\end{equation}
where the functionals $R(u)$ and $F(u;f)$, commonly referred to as the \emph{regularization} and the \emph{fidelity} term,
encode prior information on the clean image $u$ and the observation model, respectively,
with the so-called regularization parameter $\lambda > 0$ controlling the trade-off between the two terms.
In particular, the functional form of the fidelity term is strictly connected to the characteristics of the noise corruption.
A classical choice for the fidelity measures the data fitting in terms of the 
$\ell_2$-norm, in formulas:
\begin{equation}
F(u;f) \,\;{:=}\;\,  \, \| u - f \|_2^2.
\label{eq:Lq}
\end{equation}

For what regards the regularization term $R(u)$ in (\ref{eq:GVM}), a very popular choice
is represented by the Total Variation, presented in the following two forms:
\begin{eqnarray}
\text{[Isotropic TV]} & R_{\text{iso}}(u) := &  
\| Du \|_{2,1} =  \| \sqrt{
 (D_{X} u)^2 + (D_{Y} u )^2} \|_1 
 \label{eq:isotropic}\\
\nonumber\\
%\end{array}
%\end{equation}
%\begin{equation}
%\begin{array}{crl}
\text{[Anisotropic TV]} & R_{\text{ani}}(u) := &  
\| Du \|_{1} =  \|D_{X}u\|_1 + \|D_{Y}u\|_1
\label{eq:anisotropic}
\end{eqnarray}
where $D_x, D_y$ denote the horizontal and vertical operators, respectively, and $D = (D_x, D_y)$ is the gradient operator $\nabla$ in the discrete setting.

By substituting the TV regularizer $TV(u):=R_{iso}(u)$ in (\ref{eq:isotropic}) or $TV(u):=R_{ani}(u)$ in (\ref{eq:anisotropic})  and the fidelity term (\ref{eq:Lq}) for $R$ and $F$ in (\ref{eq:GVM}),
respectively, one obtains the so-called TV-L$_2$  - restoration model, originally introduced in \cite{ROF92}. In formulas:
\begin{equation}
u^* \:\;{\leftarrow}\;\: \arg \min_{u \in \mathbb{R}^N}
\left\{ \,
J(u)=\mathrm{TV}(u) \,\;{+}\;\, \lambda \, \| u - f \|_2^2
\, \right\} \, .
\label{eq:TVLq}
\end{equation}

\subsection{Median formula for TV}\label{MedForm}
 Li and Osher in \cite{OsherTV} proposed an efficient and highly parallelizable method for solving TV model (\ref{eq:TVLq}) with  anisotropic TV regularization \eqref{eq:anisotropic}. They proposed to solve iteratively $N$ one-dimensional optimization problems to obtain an accurate solution of the $N$-dimensional optimization problem (\ref{eq:TVLq}).
 In particular, for each pixel $u\in\mathbb{R}$, we consider the local minimization problem:
\begin{equation} \label{eq:E}
u^* = \arg\min_{u \in \mathbb{R}} \left\{ E(u) := \sum_{i = 1}^{\kappa}w_{i}\lvert u - u_{i} \rvert + F(u)\right\}
\end{equation}
where  $F(u) = \lambda(f - u)^{2}$, $u^*, f \in \mathbb{R}$ are respectively the denoised and noisy versions of the same pixel, while $u_i$ belongs to the set of $\kappa$  neighboring pixels and $w_i\ge0$ are given weights.
In \cite{OsherTV}, a simple method for computing $u^*$ in (\ref{eq:E}) is derived and here reported for self-consistency.

\begin{theorem}\label{median theorem}
Supposing the $w_i$ are non-negative and the $u_i$ are sorted as $u_1 \leq u_2 \leq ... \leq u_{\kappa}$, the function $F$ is strictly convex and differentiable and $F'$ is bijective; then the minimizer of (\ref{eq:E}) is a median:
\begin{equation} \label{eq:median}
u^* = \textup{median}\{u_1, u_2, ... , u_{\kappa}, p_0, p_1, ... , p_{\kappa} \}
\end{equation}
where $p_i = (F')^{-1}(W_i)$ and 
\begin{equation} \label{eq:W}
W_i = -\sum_{j = 1}^{i}w_j +\sum_{j = i + 1}^{\kappa}w_j,\,\,\,i = 0, 1, \ldots, {\kappa}.
\end{equation}
\end{theorem}
In our formulation, the neighborhood of the current pixel $u$ are simply $u_u, u_d, u_l, u_r$, the vertical and horizontal direct neighbors pixels, respectively. 
The adopted 4-neighbors strategy allows us to apply the median formula in parallel on multiple pixels at one time using a proper configuration, since each pixel is directly affected only by its 4 neighbors.

The fidelity is defined as  $F(u) = \lambda(f - u)^{2}$, and consequently $$ F'(u) = -2\lambda(f - u) \Rightarrow  (F')^{-1}(W) = f + \frac{W}{2\lambda},$$
where $W$ is a sum of weights previously defined in \eqref{eq:W}.
The denoised pixel $u^*$ is then obtained as the median value: 
\begin{equation}
    u^* = \text{median}\{u_l, u_r, u_u, u_d, p_0, p_1, p_2, p_3, p_4\},
\end{equation}
where the p-values $p_i$ are calculated following Theorem  \ref{median theorem}, with $w_i=1$ and $\kappa = 4$:
\begin{equation}
\begin{array}{lll}
W_0 = &4	\Rightarrow p_0 = f + \frac{4}{2\lambda}	\Rightarrow & p_0 = f + \frac{2}{\lambda}
\\
W_1 = &2	\Rightarrow p_1 = f + \frac{2}{2\lambda}	\Rightarrow & p_1 = f + \frac{1}{\lambda}
\\
W_2 = &0	\Rightarrow p_2 = f + \frac{0}{2\lambda}	\Rightarrow & p_2 = f
\\
W_3 = &-2	\Rightarrow p_3 = f + \frac{-2}{2\lambda}	\Rightarrow & p_3 = f - \frac{1}{\lambda}
\\
W_4 =& -4	\Rightarrow  p_4 = f + \frac{-4}{2\lambda}	\Rightarrow & p_4 = f - \frac{2}{\lambda}
\end{array}
\label{eq:Wi}
\end{equation}

The image denoising problem can hence be solved by iteratively computing (\ref{eq:median}) pixel-by-pixel over the whole image until convergence, which is guaranteed by the following result. 

\begin{theorem}\label{convergence theorem}
The algorithm defined by repeatedly applying (\ref{eq:median}) at the $j$th pixel, converges, i.e. $u_j^{(k+1)} = \arg\min_{u_j\in\mathbb{R}}E^{(k)}(u_j)$, hence 
$$
u^{(k)} \Rightarrow \arg\min_{u \in {\mathbb R}^N} J(u).
$$
\end{theorem}
This means that, after  $k$ iterations, with $k \rightarrow \infty$, we obtain the minimizer of problem (\ref{eq:TVLq}). 

For the numerical implementation the stopping criterion considered is computed as follows: given a small tolerance value $\epsilon$, process stops when, at iteration $k$
\begin{equation}\label{eq:convergence}
\frac{\|u^{(k-1)} - u^{(k)}\|_2}{\|u^{(k-1)}\|_2} \leq \epsilon
\end{equation}
where $u^{(k-1)}$ and $u^{(k)}$ are consecutive processed images.

The resulting algorithm is described in Algorithm \ref{alg:ATV}.

\begin{center}
%\begin{minipage}{0.48\textwidth}
\begin{algorithm}[H]
 \textbf{Input:}  $f \in \mathbb{R}^N$, $\lambda>0$ \; \\
 \textbf{Output:} $u^*\in \mathbb{R}^N$   \\
\textbf{Initialize}   $\mbox{image}^{(0)} = f, k=0$\\
\textbf{while(!convergence)} \\
 \begin{tabular}{ll}
%    &Vectorize $y$\\
%  & $a_0 = 0$ & \;\\
   &\textbf{For each} pixel \textbf{in} $\mbox{image}^{(k)}$ \textbf{do:}   \\
   &\quad compute $W\in \mathbb{R}^5$ as in \eqref{eq:Wi} \\
   &\quad set $N_{pixel} =$ get\_neighbors(pixel) 
   $\quad \%N_{pixel}=(u_l,u_r,u_d,u_u)$ \\
 & $\quad v \leftarrow$ sort($N_{pixel}$) in ascending order \\
  & $\quad$ compute $p \in \mathbb{R}^{5}$, with $p_j= $ pixel $+\frac{1}{\lambda} W_j, \quad j=0,..,4$\\
  &  
  $\quad$ pixel$^* \leftarrow \mbox{median}(v_0,v_1,v_2,v_3,p_0,p_1,p_2,p_3,p_4)$ \\
  & $\quad$ tmp\_image $\leftarrow$ pixel$^*$\\
  & \textbf{end}  
 \end{tabular}\\
 $\mbox{image}^{(k+1)}$ =tmp\_image\\
 $k=k+1$\\
 \textbf{end} \\
 $u^*= \mbox{image}^{(k+1)}$\\
 \caption{Anisotropic TV algorithm}
 \label{alg:ATV}
\end{algorithm}
%\end{minipage}
\end{center}

\lstset{% general command to set parameter(s)
	basicstyle=\small,
	title=Anisotropic TV algorithm,
	frame=single,
	tabsize=3
	}

\section{Quantum Image Processing}
\label{sec:sec3}

%Quantum Image Processing (QIP) refers to the application of quantum computing mechanisms to image processing. 
QIP aims to design quantum algorithms that, once constructed  a quantum state which encodes an image, implement image processing techniques in a quantum environment. 
QIP is a novel topic in quantum computing and many issues are far from being solved, as explained in \cite{review4}. A central issue regards the image encoding in a quantum environment, which will be discussed in Section 
\ref{sec:qir}. Another limiting issue is that, nowadays, QIP represents a demanding branch of quantum computing, as it needs a lot of resources that are far from being offered by current or medium term devices. At the moment, researchers are proposing many different approaches to the problem, even though only a limited number of these are commonly accepted and used \cite{review1,review2,review3}. 
In order to tackle the memory restrictions, we subdivide the image into a set of sub-images as described in Section \ref{sec:patches}. 

\begin{figure}[h]
\centering
\includegraphics[width=.98\linewidth]{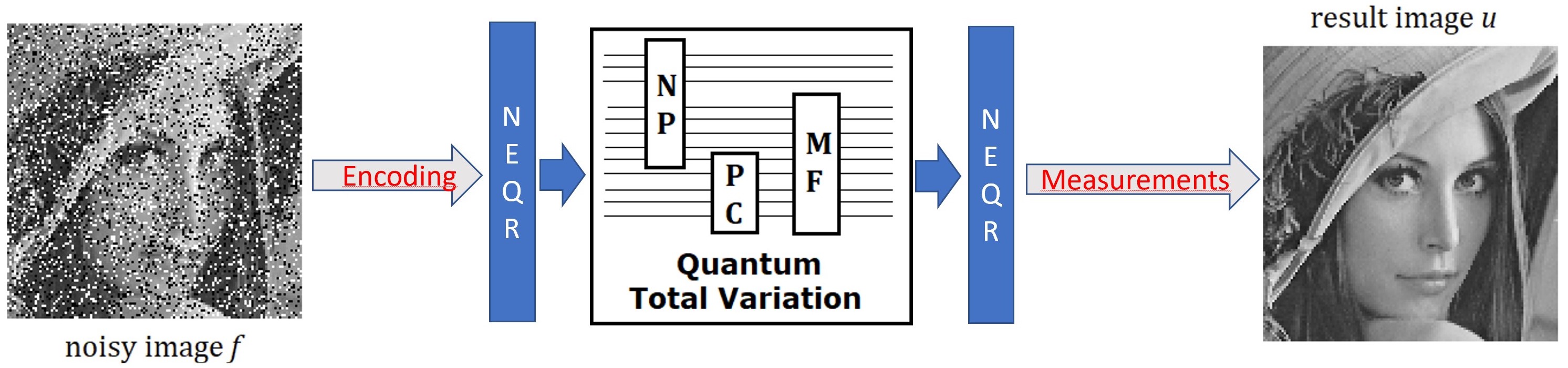}
\caption{Scheme illustrating Quantum TV (QTV) denoising process.}
\label{fig:process}
\end{figure}

The denoising QIP here proposed consists of three steps, illustrated in Fig.\ref{fig:process}.
The image is first encoded in the quantum environment, then it is processed by quantum circuits to perform the denoising task, and finally the denoised image is measured to convert it in a classical image format.
The used measurement process is detailed in Section \ref{sec:measure}. 

\subsection{Quantum Image Representation}\label{sec:qir}
A quantum image encoding is defined Quantum Image Representation (QIR). Unlike classical image processing, where a set of well-known standard formats are available and well-assessed, in QIP many encoding QIR techniques were proposed and tested \cite{QM_orig}, but on the other hand none nowadays has distinguished itself as standard.
The most used QIR technique is the 
 Novel Enhanced Quantum Representation (NEQR).
 
 This kind of representation needs to encode the following image's data:
\begin{itemize}
\item
\textbf{Pixel coordinates}, encoded by qubits $\ket{XY}$. An image of dimension $D_{X} \times D_{Y}$, usually needs to use $d_{X} = \lceil \log_{2}D_{X} \rceil$ qubits for horizontal coordinates and 
$d_{Y} = \lceil \log_{2}D_{Y} \rceil$ qubits for vertical ones. In this way $ \ket{XY} = \ket{X_0 X_1 ...  X_{d_x-1} Y_0 Y_1 ...  Y_{d_y-1}}$.
\item
\textbf{Pixel value}, encoded by one or more qubits  $\ket{C}$. It uses $q$ qubits for representing $N_q=2^q$  possible values in a binary encoding. 
\end{itemize}

Without loss of generality, we consider grayscale images $I$ with a square $2^n \times 2^n$ domain and values in the range $[0,2^q-1]$. In this way we consider $\ket{XY}$ with $2n$ qubits, where $n = \lceil\log_{2}D\rceil$, and $\ket{C}$ using $q=\lceil\log_{2}N_q\rceil$ qubits to represent $2^q$ possible values. 
%
%\begin{center}
%\begin{quantikz}
%\lstick{$\ket{0}^{\otimes q}$} & \gate[wires=3]{\text{QIR}} & \rstick{\ket{C}}\qw\\
%\lstick{$\ket{0}^{\otimes n}$} &  & \rstick{\ket{Y}}\qw\\
%\lstick{$\ket{0}^{\otimes n}$} &  & \rstick{\ket{X}}\qw
%\end{quantikz}
%\end{center}
%
Therefore, a grayscale image $I$ 
 %with size $2^n \times 2^n$ and gray range from $0$ to $2q - 1$ (usually $2q - 1 = 255$)
 is encoded in the following quantum state of $2n + q$ qubits:
\begin{equation}
\label{eq:NEQR}
%\begin{aligned}
\ket{I(n)} = \frac{1}{2^n}\sum_{X=0}^{2^{n}-1}\sum_{Y=0}^{2^{n}-1}\ket{C_{XY}} \otimes \ket{XY},
%\end{aligned}
\end{equation}
where
\begin{equation}\label{eq:NEQR}
\begin{aligned}
\ket{C_{XY}} = \ket{C_{XY}^{q-1} \ldots C_{XY}^2 C_{XY}^1 C_{XY}^0} \in \{0,1,\ldots,2^{q}-1\}\\
\end{aligned}
\end{equation}
and each $\ket{C_{XY}^i}$ qubit is $\ket{0}$ or $\ket{1}$. 
%With this binary encoding we're able to use arithmetic operations on color data in a simpler way than using FRQI. As NEQR uses basis states for colors, it is way less prone to noise errors.\\
%
\begin{figure}[t]
	\centering
	\begin{tabular}{cc}
	\includegraphics[width=3.0cm]{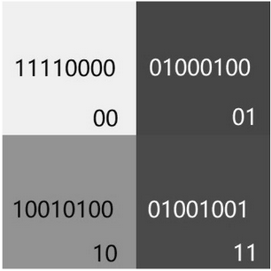} & \includegraphics[width=4.5cm]{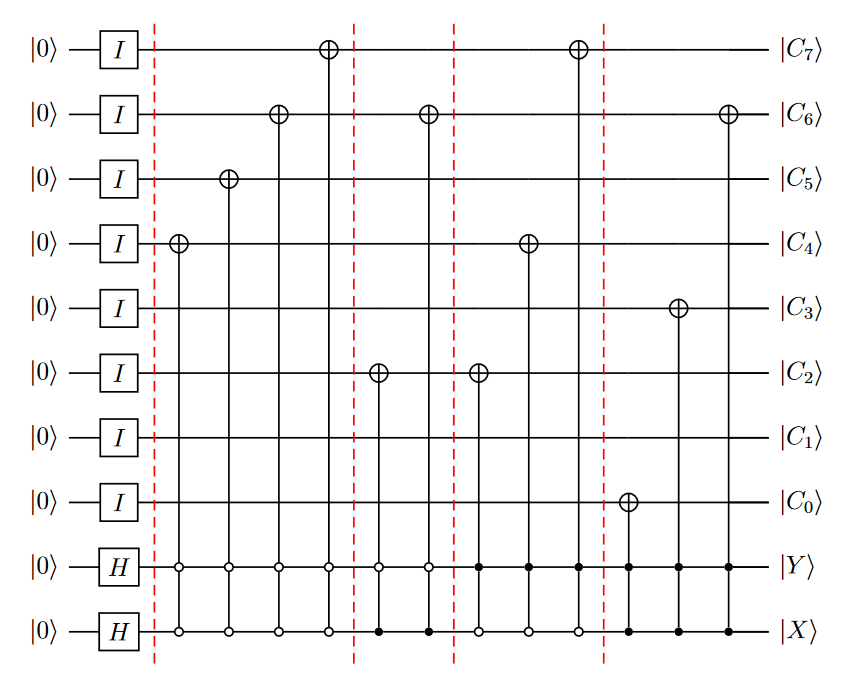}\\
	\end{tabular}
	\caption{NEQR representation of an image: (left) the $2\times 2$ image; (right) NEQR circuit, encoding pixels 00, 01, 10, 11 (left to right). Figure on the left from \cite{review1}.}
	\label{fig:NEQR_test}
\end{figure}
A simple example of NEQR representation for a $2 \times 2$ grayscale image is illustrated in Fig. \ref{fig:NEQR_test}(left),  where each corner number denotes pixel's coordinates, while centered value indicates pixel's intensity \cite{review1}. The quantum wave function for this image in NEQR is hence the following:
\begin{equation}\label{eq:NEQR_test}
\begin{aligned}
\ket{I} =&\frac{1}{2}(\ket{C_{00}}\otimes\ket{00} + \ket{C_{01}}\otimes\ket{01} + \ket{C_{10}}\otimes\ket{10} \ket{C_{11}}\otimes\ket{11})\\
=&\frac{1}{2}(\ket{11110000}\otimes\ket{00} + \ket{01000100}\otimes\ket{01} +\\
& + \ket{10010100}\otimes\ket{10} + \ket{01001001}\otimes\ket{11})
\end{aligned}
\end{equation}

This is obtained by putting the coordinate qubits $\ket{X}$ and $\ket{Y}$ in a superposition state using Hadamard gates, then entangle them with qubits $\ket{C_{XY}}$ using a series of CNOT gates. The resulting NEQR circuit is illustrated in Fig. \ref{fig:NEQR_test}(right) for the image in Fig. \ref{fig:NEQR_test}(left).\\

\smallskip

NEQR is a very versatile representation for image computation. However, it presents some drawbacks.
%\begin{itemize}
%\item
As pixels are encoded one by one, large images produce long NEQR circuits; this means that circuit depth grows linearly with image size.
%\item
As the image size grows, the number of coordinate qubits become larger and the control qubits needed for encoding quickly become more than two. A Toffoli gate with more than two control qubits is defined Multiple CNOT gate, or just MCX: this operator is decomposed before execution, using many Toffoli gates and some auxiliary qubits called \emph{ancilla}. From an efficiency point of view, more coordinate qubits means larger MCX to be used, which leads to a larger amount of gates.

\subsection{Image pre-processing}
\label{sec:patches}
In order to reduce the memory usage, we split a pre-padded image (one-pixel bordered) into many smaller overlapped patches, $4 \times 4$ sub-images. Once extracted, a patch of pixels will be processed by the quantum TV algorithm.  At the end of each iteration, will be necessary to reassemble the resulting image from the output patches.

\begin{figure}[b!]
\centering
\includegraphics[width=.90\linewidth]{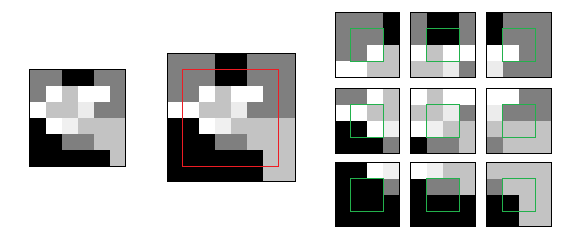}
\caption{Image pre-processing: image (left), padded image (center) and extracted patches (right).}
\label{fig:preprocess}
\end{figure}

Fig. \ref{fig:preprocess} illustrates the image pre-processing procedure for a sample image: red square frames original image, while green squares in patches highlight the processed pixels.

In order to speed up quantum circuits generation, we have subdivided workload using multi-threading: each thread is tasked to assemble a QTV for each image patch, using pre-assembled circuits and generating remaining ones. This approach considerably accelerated the execution process for what regards the generation phase.\\

\subsection{Image measurement}
\label{sec:measure}
The image extraction from the QIR format is a not-trivial process and its performance depends on the quantum representation used in the algorithm. 

A quantum representation collects all image's data in a single quantum state.
Due to the nature of this particular quantum state, the extraction is not a deterministic process: for each measurement, one of the possible pixel coordinate-value association is randomly obtained as outcome from the \emph{collapse} of the quantum state. For a complete recovery of an image, it is necessary to execute the same algorithm many times.

The image measurement is the last step in Fig.\ref{fig:process}.

%\end{itemize}
%
%\section{Algorithm application}
%A QIP algorithm is applied to a quantum image representation. In order to be used, these algorithms have to be specific for the used representation, as color encoding modifies how values are processed throughout the circuit.
%
%A QIP algorithm is a sub-circuit applied to qubits involved in image representation. It can use a number of auxiliary qubits that will later be discarded. To perform complex image processing, a series of sub-circuits are used, one for each transformation phase. The general structure of a QIP process is the following:
%
%\begin{center}
%\begin{quantikz}
%\lstick{$\ket{0}^{\otimes q}$} & \gate[wires=3]{\text{QIR}}& \gate[wires=3]{\text{$U_1$}} 	& \gate[wires=3]{\text{$U_2$}}	& \gate[wires=3]{\text{$U_3$}}	&\rstick{\ket{C^{'}}}\qw\\
%\lstick{$\ket{0}^{\otimes n}$} &  				  & 					&					&					&\rstick{\ket{Y^{'}}}\qw\\
%\lstick{$\ket{0}^{\otimes n}$} &  				  &					&					&					&\rstick{\ket{X^{'}}}\qw
%\end{quantikz}
%\end{center}
%where $U_1, U_2\, \text{and} \,U_3$ are circuits implementing image processing algorithms. We can have for example:
%\begin{itemize}
%\item
%$\text{QIR}$ encodes a noisy image;
%\item
%$U_1$ is a denoising alg. implementation, which \emph{cleans} input image;
%\item
%$U_2$ is an edge detection alg.;
%\item
%$U_3$ is a feature extraction alg.
%\end{itemize}
%Eventually the last sub-circuit could be different kind of algorithm, not QIP related, but an efficient implementation for processing input as a dataset more than as an image.

\section{Quantum TV Filter}
\label{sec:sec4}

In this section we introduce the Quantum TV Filter, named QTV, a quantum implementation of the TV regularization algorithm described in Section \ref{sec:sec1} applied to qubits involved in the NEQR quantum image representation.
This work extends and improves the work in \cite{QM} where the authors proposed a quantum solution  for implementing a simple median filter for image processing.

\begin{figure}[t]
	\centering
	\begin{tabular}{cc}
	\includegraphics[width=4.0cm]{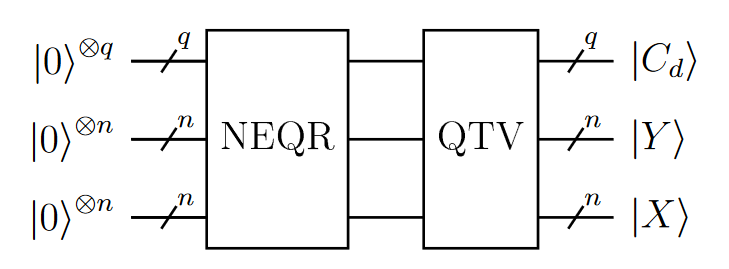} & \includegraphics[width=4.5cm]{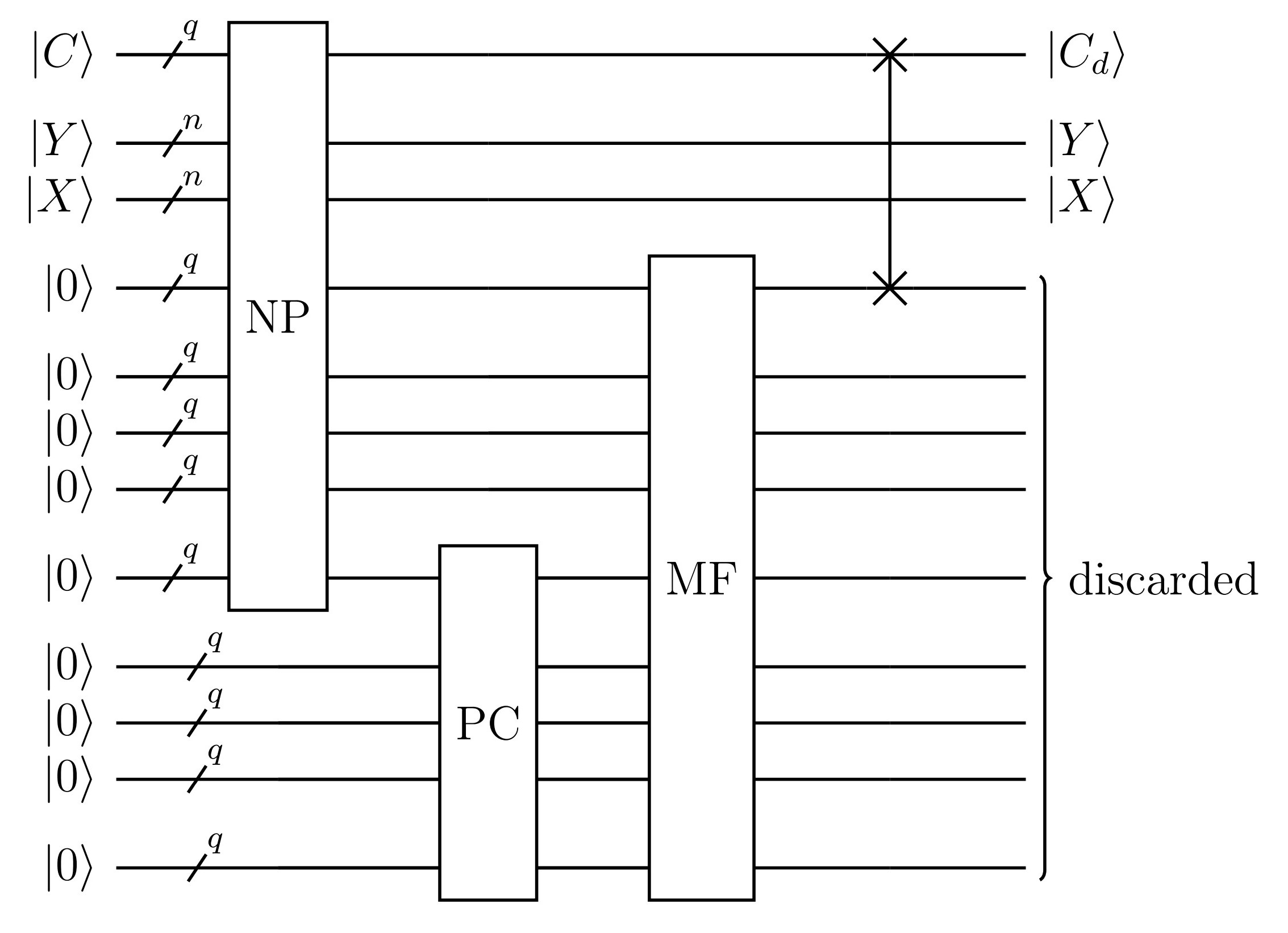}\\
	\end{tabular}
	\caption{Quantum TV general scheme (left); a detail of the QTV sub-circuits (right)}
	\label{fig:QMF}
\end{figure}

The proposed circuit processes an image input in NEQR representation and provides a denoised image in output in the same NEQR form, according to the scheme in Fig. \ref{fig:QMF}(left).

According to the TV algorithm presented in Section \ref{sec:sec1} we have to iterate a core process for each pixel of the input image. Considering a four-pixel neighborhood configuration, this process is composed by three steps defined as three different quantum operators acting for each pixel:
\begin{enumerate}
\item
\textbf{Neighborhood Preparation} (NP): collect neighboring pixels  and extract their values $u_u,u_d,u_l,u_r$;
\item
\textbf{P-values Computation} (PC): compute weighted values $p_0,p_1,p_2,p_3,p_4$;
\item
\textbf{Median Function} (MF): extract median value from set $\{u_u,u_d,u_l,u_r,p_0,p_1,p_2,p_3,p_4\}$
\end{enumerate}
The quantum operators are assembled into the QTV structure illustrated in Fig. \ref{fig:QMF}(right).

In the following, we will describe in detail the three operands which characterize the Quantum TV. Each one is composed of many sub-circuits, or modules.

\begin{figure}[t]
	\centering
	\includegraphics[width=12.0cm]{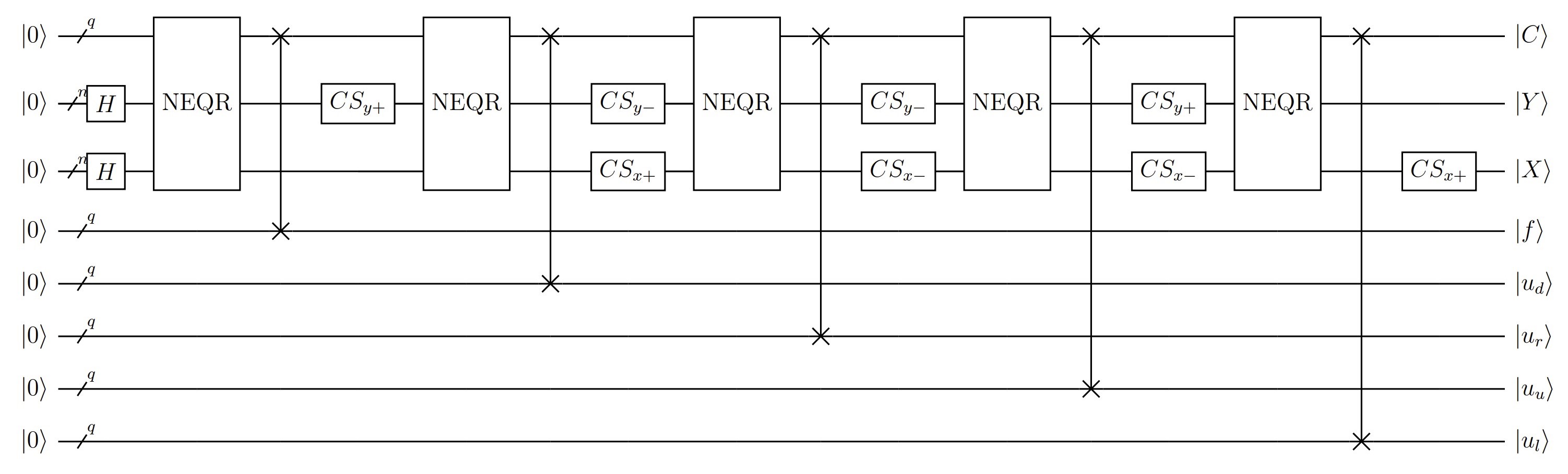}  \caption{Neighborhood Preparation module}
	\label{fig:NP}
\end{figure}

\subsection{Neighborhood Preparation}
This operand is in charge of extract neighboring pixels from NEQR representation. At this aim, we used Cycle-Shift (CS) module to change a coordinate register's value, allowing us to shift an image up, down, left or right; see Appendix A for details.
%CS is essentially a modulo-2 adder or subtractor, defined as CS+ and CS-, respectively, illustrated in the Appendix.

The structure of the NP operand, which output is a quantum superposition state of the  central $f$, up $u_u$, down $u_d$, left $u_l$ and right $u_r$ pixel values, is illustrated in Fig. \ref{fig:NP}.
Specifically, once obtained a pixel value from NEQR, we use CS to shift coordinate values, then we re-apply NEQR to gather a new pixel value corresponding to the new position data.
With the exception of the first NEQR for image encoding, we avoid the usage of H-gates applied to coordinate registers, as we want to extract a specific (and not random) color outcome.

The output of the NP operand is a quantum state obtained starting from $\ket{\psi} = \ket{0}^{\otimes(q+2n+5q)}$, which reads as
\begin{equation}
\ket{\psi_{NC}}=\frac{1}{2^n}\sum_{X=0}^{2^{n}-1}\sum_{Y=0}^{2^{n}-1}\ket{u_l}\ket{u_u}\ket{u_r}\ket{u_d}\ket{f}\ket{0}\ket{X}\ket{Y}	
\end{equation}
where each qubit in the $\ket{C}$  register is  reset to the $\ket{0}$ state. 

More in detail, following the scheme in Fig.\ref{fig:NP}, we have

$$
\begin{array}{rll}
\text{NEQR}\cdot\text{H}_{XY}\ket{\psi}=		&\frac{1}{2^n}(\sum_{X=0}^{2^{n}-1}\sum_{Y=0}^{2^{n}-1}\ket{f}\ket{X}\ket{Y})\otimes\ket{0}^{\otimes5}									&=\ket{\psi_1}	\\
\text{SWAP}\ket{\psi_1}=				&\frac{1}{2^n}(\sum_{X=0}^{2^{n}-1}\sum_{Y=0}^{2^{n}-1}\ket{f}\ket{0}\ket{X}\ket{Y})\otimes\ket{0}^{\otimes4}								&=\ket{\psi_2}	\\
\text{CS}_{y+}\ket{\psi_2}=				&\frac{1}{2^n}(\sum_{X=0}^{2^{n}-1}\sum_{Y=0}^{2^{n}-1}\ket{f}\ket{0}\ket{X}\ket{Y+1})\otimes\ket{0}^{\otimes4}								&=\ket{\psi_3}	\\
\text{NEQR}\ket{\psi_3}=				&\frac{1}{2^n}(\sum_{X=0}^{2^{n}-1}\sum_{Y=0}^{2^{n}-1}\ket{f}\ket{u_d}\ket{X}\ket{Y+1})\otimes\ket{0}^{\otimes4}							&=\ket{\psi_4}	\\
\text{SWAP}\ket{\psi_4}=				&\frac{1}{2^n}(\sum_{X=0}^{2^{n}-1}\sum_{Y=0}^{2^{n}-1}\ket{u_d}\ket{f}\ket{0}\ket{X}\ket{Y+1})\otimes\ket{0}^{\otimes3}					&=\ket{\psi_5}	\\
\text{CS}_{y-}\text{CS}_{x+}\ket{\psi_4}=		&\frac{1}{2^n}(\sum_{X=0}^{2^{n}-1}\sum_{Y=0}^{2^{n}-1}\ket{u_d}\ket{f}\ket{0}\ket{X+1}\ket{Y})\otimes\ket{0}^{\otimes3}					&=\ket{\psi_6}	\\
\text{NEQR}\ket{\psi_6}=				&\frac{1}{2^n}(\sum_{X=0}^{2^{n}-1}\sum_{Y=0}^{2^{n}-1}\ket{u_d}\ket{f}\ket{u_r}\ket{X+1}\ket{Y})\otimes\ket{0}^{\otimes3}				&=\ket{\psi_7}	\\
\text{SWAP}\ket{\psi_7}=				&\frac{1}{2^n}(\sum_{X=0}^{2^{n}-1}\sum_{Y=0}^{2^{n}-1}\ket{u_r}\ket{u_d}\ket{f}\ket{0}\ket{X+1}\ket{Y})\otimes\ket{0}^{\otimes2}			&=\ket{\psi_8}	\\
\text{CS}_{y-}\text{CS}_{x-}\ket{\psi_8}=		&\frac{1}{2^n}(\sum_{X=0}^{2^{n}-1}\sum_{Y=0}^{2^{n}-1}\ket{u_r}\ket{u_d}\ket{f}\ket{0}\ket{X}\ket{Y-1})\otimes\ket{0}^{\otimes2}			&=\ket{\psi_9}	\\
\text{NEQR}\ket{\psi_9}=				&\frac{1}{2^n}(\sum_{X=0}^{2^{n}-1}\sum_{Y=0}^{2^{n}-1}\ket{u_r}\ket{u_d}\ket{f}\ket{u_u}\ket{X}\ket{Y-1})\otimes\ket{0}^{\otimes2}		&=\ket{\psi_{10}}	\\
\text{SWAP}\ket{\psi_{10}}=				&\frac{1}{2^n}(\sum_{X=0}^{2^{n}-1}\sum_{Y=0}^{2^{n}-1}\ket{u_u}\ket{u_r}\ket{u_d}\ket{f}\ket{0}\ket{X}\ket{Y-1})\otimes\ket{0}			&=\ket{\psi_{11}}	\\
\text{CS}_{y+}\text{CS}_{x-}\ket{\psi_{11}}=	&\frac{1}{2^n}(\sum_{X=0}^{2^{n}-1}\sum_{Y=0}^{2^{n}-1}\ket{u_u}\ket{u_r}\ket{u_d}\ket{f}\ket{0}\ket{X-1}\ket{Y})\otimes\ket{0}			&=\ket{\psi_{12}}	\\
\text{NEQR}\ket{\psi_{12}}=				&\frac{1}{2^n}(\sum_{X=0}^{2^{n}-1}\sum_{Y=0}^{2^{n}-1}\ket{u_u}\ket{u_r}\ket{u_d}\ket{f}\ket{u_l}\ket{X-1}\ket{Y})\otimes\ket{0}		&=\ket{\psi_{13}}	\\
\text{SWAP}\ket{\psi_{13}}=				&\frac{1}{2^n}\sum_{X=0}^{2^{n}-1}\sum_{Y=0}^{2^{n}-1}\ket{u_l}\ket{u_u}\ket{u_r}\ket{u_d}\ket{f}\ket{0}\ket{X-1}\ket{Y}			&=\ket{\psi_{14}}	\\
\text{CS}_{x+}\ket{\psi_{14}}=				&\frac{1}{2^n}\sum_{X=0}^{2^{n}-1}\sum_{Y=0}^{2^{n}-1}\ket{u_l}\ket{u_u}\ket{u_r}\ket{u_d}\ket{f}\ket{0}\ket{X}\ket{Y}			&=\ket{\psi_{NC}}	\\	
\end{array}
$$
\begin{figure}[h]
	\centering
	\includegraphics[width=5.0cm]{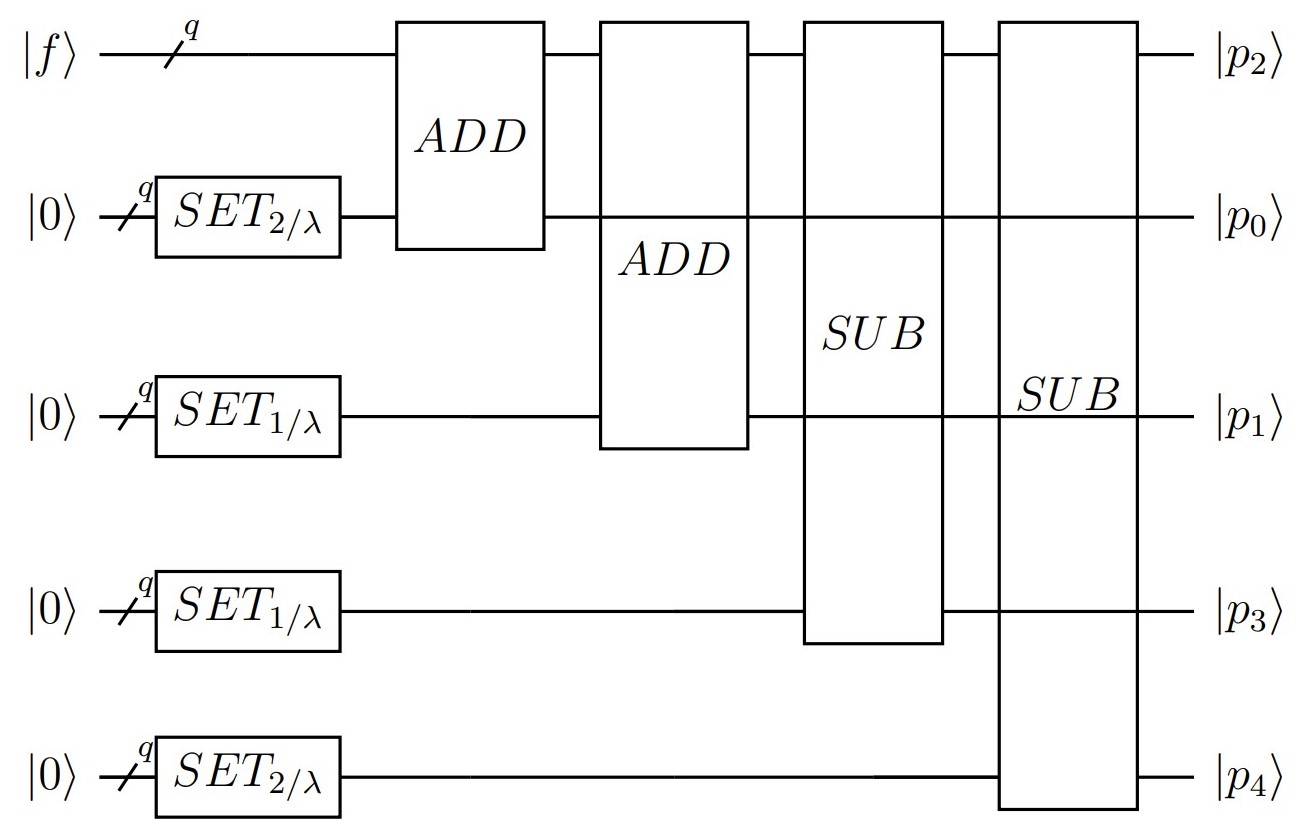}  \caption{P-values computation module}
	\label{fig:PC}
\end{figure}

\subsection{P-values computation}

Starting from the obtained neighborhood values, we compute the $p$-values according to relations \eqref{eq:Wi}. 
This reduces to  adding some constant values to $f$, the central pixel.

However the QTV algorithm only works in unsigned integer
arithmetic, while the TV algorithm works with floating point numbers, thus getting more accurate and precise results. Therefore,  instead of the $p$-values 
$p_i = f + \frac{W_i}{2\lambda},$
we force approximated rounded values 
\begin{equation}p_i = f + round(\frac{W_i}{2\lambda}).
\end{equation}
%where we considered $1/\lambda$ as regularization parameter and the round to the nearest integer.
The $p$-values are then given by:
\begin{equation}
\begin{array}{lll}
W_0 = 4,	& p_0 = & f + round(2/\lambda)\\
W_1 = 2,	& p_1 = & f + round(1/\lambda)\\
W_2 = 0,	& p_2 = & f\\
W_3 = -2,	& p_3 = & f - round(1/\lambda)\\
W_4 = -4,	& p_4 = & f - round(2/\lambda).
\end{array}
\end{equation}

The final design of the P-values Computation module is hence illustrated in Fig. \ref{fig:PC}, and it is composed by three sub-circuits for setting, adding and subtracting the mentioned constants.  
In particular:
\begin{itemize}
    \item SETTER module: assign the round to the nearest integer of a given value to a register, which is used to encode our constants;
    \item ADDER module: add two values encoded in two quantum registers. We refer to the Appendix A for more details; 
    \item SUBTRACTOR module: subtract two values by using an adder module, according to $\alpha - \beta \Rightarrow \overline{ \overline{\alpha} + \beta}.$
    \end{itemize}

\subsection{Median function}

\begin{figure}[b!]
\centering
\includegraphics[width=.98\linewidth]{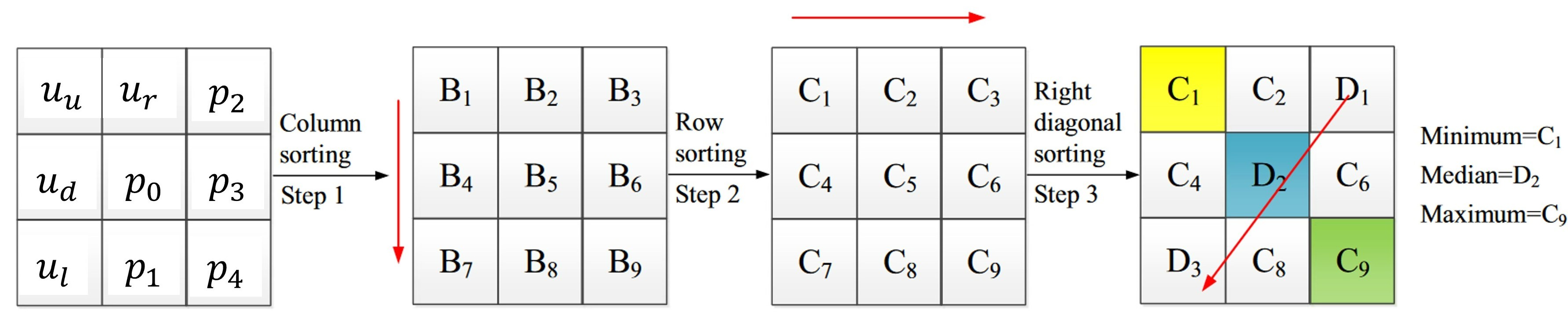}
\caption{The basic sorting strategy.}
\label{fig:sort}
\end{figure}

To determine a median from a set of values, we need to sort them. 
The sorting strategy here adopted follows the proposal in \cite{QM} for a set of 9 sortable elements.
If we re-arrange these values in a matrix form, then we simply need to follow these three steps in pipeline:
sort each column,
sort each row, and sort the right diagonal.
This will guarantee us to store the median value in the central quantum register.
 The strategy is illustrated in Fig. \ref{fig:sort}.

\begin{figure}[b!]
\centering
\includegraphics[width=7cm]{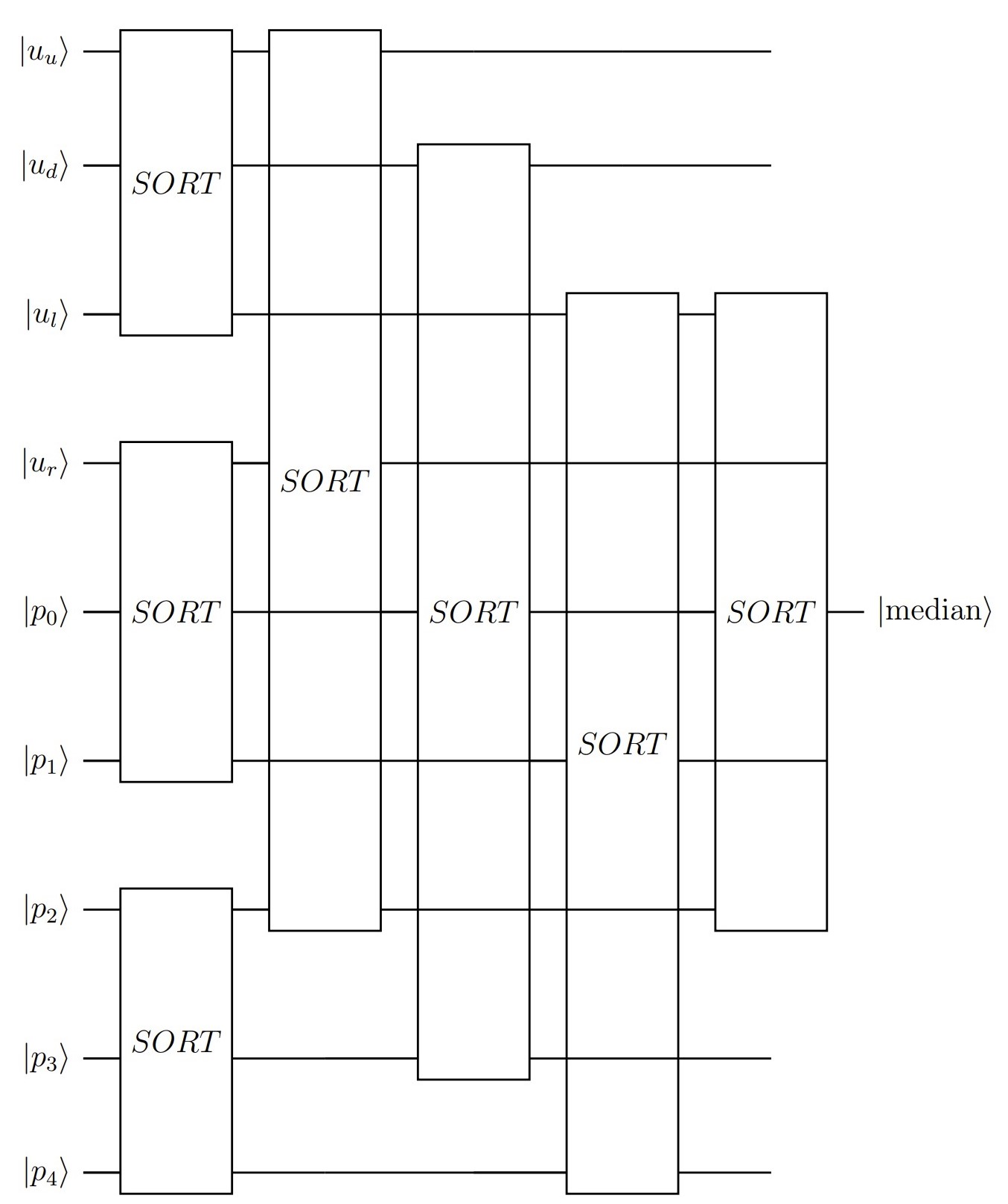},
\caption{Median function module}
\label{fig:MF}
\end{figure}

The corresponding quantum circuit is shown in Fig. \ref{fig:MF}.

\begin{figure}[b!]
\centering
\includegraphics[width=7.5cm]{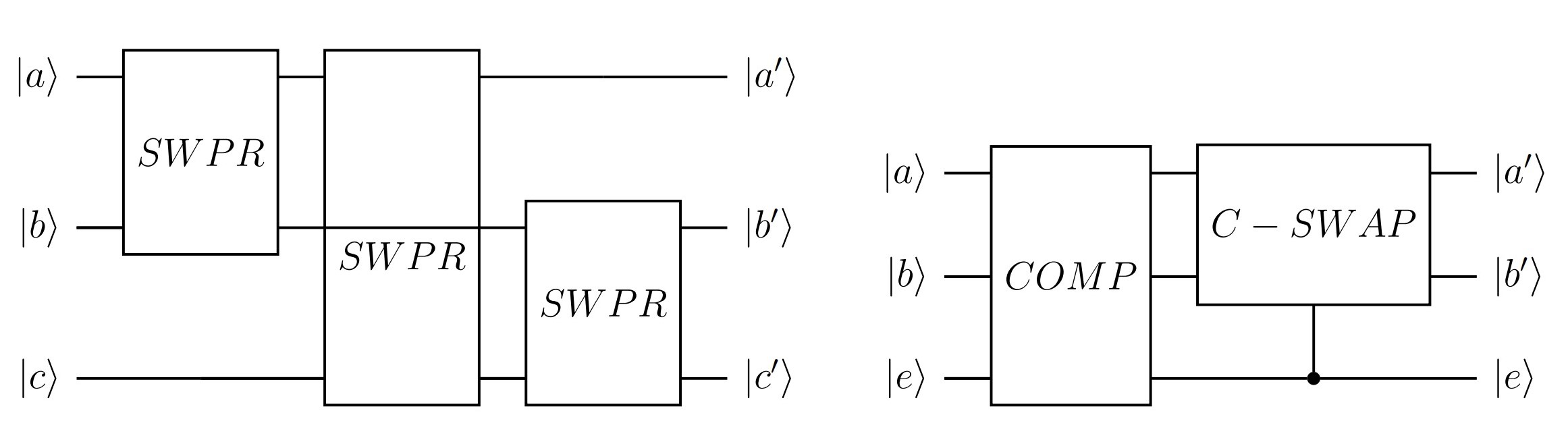},
\caption{Sort module (left); Swapper module (right)}
\label{fig:s_s}
\end{figure}

The Sort module is then the core of
Median Function module. 
Sort module has to order its three input registers. The total ordering of three elements is a trivial problem, as it is reduced to compare two positive integer values and, if not ordered, swap them. A sequence of three comparisons is necessary and sufficient to reach a correct ordering. 
A Sort module is then a sequence of three sub-circuits, named Swapper (SWPR). A Swapper module is in turn composed by two sub-circuits: a Comparator (COMP) and a Controlled Swap (C-SWAP).
Sort and Swapper modules are shown in Fig. \ref{fig:s_s}.

The Comparator module evaluates two register values $a, b$ and provides, on an auxiliary qubit $e$, the result of the comparison $a>b$, as follows
\begin{equation}
    \text{if} \quad  a \leq b \quad \text{then} \quad e = \ket{0}
\quad \text{else} \quad e = \ket{1}.
\end{equation}

This result will be next used to control a C-SWAP gate, so that if $e = \ket{1}$, then the values $a \,\text{and}\, b$ are swapped.

The Comparator module is described in Appendix A. 
The designed 
quantum circuit for the Comparator module is more efficient with respect to other proposals. For example, in case $q=8$, this circuit uses
less quantum elementary gates than other existing methods: depth = 64 with respect to the Sort module applied in \cite{QM} which has depth = 1.091.767.

\section{Circuit complexity analysis}
\label{sec:sec5}

In order to estimate a quantum circuit efficiency, we have to look at its depth, that is the longest path in it. The path length is always an integer number, representing the number of gates it has to execute in that path. At this aim, we are going to analyze each operand of Quantum TV to derive an approximate estimation of its depth.

Neighborhood Preparation is for sure the most demanding operand of all filtering algorithm, because it uses multiple instances of NEQR circuit, which length is variable according to the number of pixel to encode. Considering a NEQR implementation that commits the least amount of MCX gates and encodes $N$ pixels, its depth grows polynomially with $N$ as follows,
$$
\text{NEQR}_{\text{depth}} = N(4(2\log_2{\sqrt{N}} -1) + 8) \Rightarrow \mathcal{O}(8N\log_2{\sqrt{N}} + 4N).
$$
%\textcolor{red}{This estimation is based on a simple reasoning. When using an optimized version of NEQR, each pixel encoding uses 2 MCX gates: given $x$ control qubits, each MCX is decomposed in $2(x-1)$ Toffoli gates. The number of control qubits is equal to the amount of coordinates qubits, that is $2\log_2{\sqrt{N}}$. This means that each encoding uses at least $4(2\log_2{\sqrt{N}}-1)$ consecutive gates, but we have also to consider CNOTs used to entangle color pixels, which in the worst case are 8. Each pixel then needs to use $4(2\log_2{\sqrt{N}}-1) + 8$ gates for being encoded, thus total NEQR depth is $N(4(2\log_2{\sqrt{N}}-1) + 8) = N(8\log_2{\sqrt{N}}-4+8)=8N\log_2{\sqrt{N}}+4N$.}

Other modules involved are Swap and Cycle Shift. Swap depth depends on the number $q$ of color qubits, thus its depth is always equal to that value. Cycle Shift depth, instead, depends on coordinate register size and it is equal to $\log_2{\sqrt{N}}\cdot(\log_2{\sqrt{N}}-1)$.

Neighborhood preparation uses NEQR, SWAP and CS five times in a row. Hence the NP module overall depth is  polynomial in $N$ and we can estimate the NP depth as:
$$
\text{NP}_{\text{depth}} = 5\cdot(8N\log_2{\sqrt{N}} + 4N + q + \log_2{\sqrt{N}}\cdot\log_2{\sqrt{N}}-1))
$$

For what concerns the P-values Computation depth,  the setting phase involves four SET modules, which however are applied simultaneously, so they count as a single one. An Adder module depth instead is dependent on the $q$ value. Full-Adder is composed of $q$ Half-Adder, with constant depth  (considering also Reset gates); then it is followed by $4q + 1$ sequentially placed gates. Thus we have:
$$
\text{ADD}_{\text{depth}} = 9q+4q+1 = 15q+1 
$$
$$
\text{SUB}_{\text{depth}} = \text{ADD}_{\text{depth}}+2 = 15q+3
$$
From these  results, we can derive the P-values Computation module total depth which is polynomial in $q$:
$$
\text{PC}_{\text{depth}} =  60q+8\Rightarrow \mathcal{O}(poly(q))
$$

To compute Median Filter depth, we just need to estimate SWPR module depth and count its occurrences in the circuit. To do so, we add up Comparator and C-SWAP depths.

Comparator's depth depends on $q$, as for each color qubit it uses 6 gates. Although Toffoli gate used in this module counts as three, as two additional X-gates need to be added in order to work. This means that module depth is estimated as $8q$. C-SWAP depth is the same as SWAP. This means that Swapper's total depth is estimated as $8q + q = 9q$.

In Median Filter there are multiple occurrences of SWPR module, which can be further reduced if row and column sorting are executed simultaneously:
$$
\text{MF}_{\text{depth}} = 9\cdot9q = 81q \Rightarrow \mathcal{O}(poly(q))
$$

From this analysis we have derived that Quantum TV algorithm  implements the NP module with polynomial complexity  in the number of pixels $N$.  The PC and MF  modules instead have a  polynomial complexity in  $q=\lceil\log_{2}N_q\rceil$, i.e. the logarithm of the number of colors, hence providing an exponential speedup. 

\section{Experimental results}
\label{sec:sec6}
In this section we evaluate  the performance of the quantum TV algorithm (QTV) on denoising grayscale images, and we present some  preliminary results from the comparison with the variational anisotropic TV in Algorithm 1.

The reference images used for the test have dimension $128\times128$ pixels and grayscale values (8-bit color depth).
The discrete model of the image degradation process under noise corruptions can be written as:
\begin{equation}\label{eq:noise}
f = \mathcal{N}(\bar{u})
\end{equation}
where $\bar{u}, f \in \mathbb{R}^{N}$ represent vectorized forms of the unknown clean image and of the observed corrupted image, respectively, while $\mathcal{N}(\,\cdot\,)$ denotes the noise corruption operator, which in most cases is of random nature.

In this work we considered two important types of noise, namely the additive (zero-mean) white Gaussian noise (AWGN) and the impulsive salt and pepper noise (SPN), which models saturated or dead pixels.

Denoting by $\Omega := \{1,\ldots,N\}$ the set of all pixel positions in the images, for these two kinds of noise the general degradation model in (\ref{eq:noise}) reads as
$$
\begin{array}{ccc}
\mathrm{AWGN:} & \mathrm{SPN:} \vspace{0.1cm} \\
f_i \:\;{=}\;\: \bar{u}_i \;{+}\; n_i \;\;\: \forall \, i \in \Omega \, ; &
f_{i} \:\;{=}\;\: \left\{
\begin{array}{ll}
	\bar{u}_i \;\; & \mathrm{for} \;\;\: i \in \Omega_0 \subseteq \Omega\\
	n_i \in \{V_{min},V_{max}\}          & \mathrm{for} \;\;\: i \in {\Omega_1} :=\Omega \setminus \Omega_0.
\end{array} \right. \, 
\end{array}
$$
In case of SPN, only a subset $\Omega_1$ of the pixels is corrupted by noise, whereas the complementary subset $\Omega_0$ is noise-free. In particular, the corrupted pixels can take only the two possible extreme values $V_{min}/V_{max}$ , where in our case assume 0 and 255 values, with the same probability. The amount of noise can be measured with an \emph{error rate} computed as follows:
$$ ER_{\%} =\frac{\text{number of corrupted pixels}}{\text{number of pixels in image}} \times 100. $$

For what concerns AWGN, the additive corruptions $n_i \in \mathcal{G}(\sigma,0)$, $i \in \Omega$, represent independent realizations from the same univariate Gaussian distribution with zero mean and standard deviation $\sigma$. 

The performance has been evaluated by the Root-Mean-Square Error (RMSE) metric, defined as:
$$ RMSE(\bar{u},u):= \sqrt{\frac{\sum_{i=1}^{N}(\bar{u}_{i} - u_{i})^2}{N}}, $$
where $\bar{u}$ is the original reference image and $u$ is the denoised output image. A lower RMSE indicates a more precise reconstruction.

For the remaining part of this work, when you come across the terms classical/quantum algorithm, we are referring to their respective implementation. 
%For a better understanding, we will label results as TV and QTV for variational and quantum algorithm respectively.

The selection of the regularization parameter $\lambda$, which exerts a crucial effect on the solution, has been carried out, for each test, by running the TV algorithm for a range of $\lambda$ values in order to select by the trial-error strategy the optimal regularization parameter. 
Then the estimated selected optimal $\lambda$ has been used in the quantum TV algorithm in order to compare the results of the two algorithms with the same optimal parameter.

A fundamental concept we must keep in mind when analyzing the obtained results, is that  classical TV denoising uses \underline{floating point} numbers, usually along with a value normalization, to be as accurate as possible. This leads to a more precise outcome and a \emph{finer} quantization of the output image. On the other hand, our quantum computation uses \underline{integer} numbers, so an approximation had to be applied (as previously described in Section \ref{sec:sec4}). This difference has an impact on the output images, which present \emph{coarser} improvements than the classical ones.

For testing purpose, we created an implementation of Quantum Median Filter using Qiskit, a Python library for quantum computing simulation \cite{Qiskit}.
The simulations of the quantum algorithm has ran on the Galileo100 supercomputer (CINECA), with the following cluster configuration:
	\begin{itemize}
	\item Nodes: 348 standard nodes
	\item Processors: 2xCPU x86 Intel Xeon Platinum 8276-8276L (2.4Ghz)
	\item Cores: 16704 (48 cores/node)
	\item RAM: 384GB
	\end{itemize}

\subsection{Example 1: AWGN denoising}

\begin{figure}[h]
\centering
	\begin{tabular}{ccc}
	 \textbf{Noisy} & \textbf{TV} &\textbf{QTV}\\ \hline \\
	\includegraphics[width=.245\linewidth]{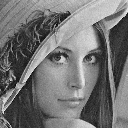} &
	\includegraphics[width=.245\linewidth]{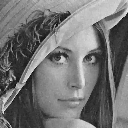} &
	\includegraphics[width=.245\linewidth]{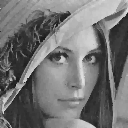} \\
	$\sigma$ = 5 & RMSE = 4.30 & RMSE = 7.21\\ 
	\includegraphics[width=.245\linewidth]{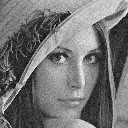} &
	\includegraphics[width=.245\linewidth]{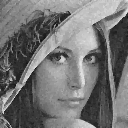} &
	\includegraphics[width=.245\linewidth]{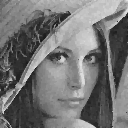} \\
	$\sigma$ = 10 & RMSE = 7.41 & RMSE = 8.80\\ 
	\includegraphics[width=.245\linewidth]{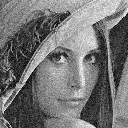} &
	\includegraphics[width=.245\linewidth]{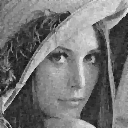} &
	\includegraphics[width=.245\linewidth]{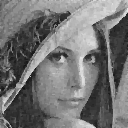} \\
		$\sigma$ = 15 & RMSE = 9.94 & RMSE = 10.49\\
	\end{tabular}
	\caption{Lena denoising results for AWGN}
	\label{fig:lena_g}
\end{figure}

\begin{figure}[h]
\centering
	\begin{tabular}{ccc}
	\textbf{Noisy} & \textbf{TV} &\textbf{QTV}\\ \hline \\
	\includegraphics[width=.245\linewidth]{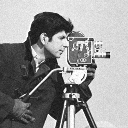} &
	\includegraphics[width=.245\linewidth]{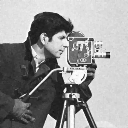} &
	\includegraphics[width=.245\linewidth]{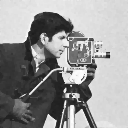} \\
		$\sigma$ = 5 & RMSE = 3.86 & RMSE = 5.99\\
	\includegraphics[width=.245\linewidth]{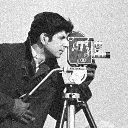} &
	\includegraphics[width=.245\linewidth]{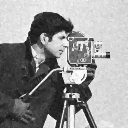} &
	\includegraphics[width=.245\linewidth]{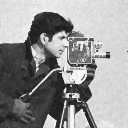} \\
	$\sigma$ = 10 & RMSE = 6.87 & RMSE = 8.14\\ 
	\includegraphics[width=.245\linewidth]{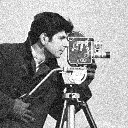} &
	\includegraphics[width=.245\linewidth]{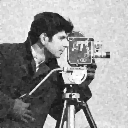} &
	\includegraphics[width=.245\linewidth]{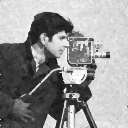} \\
	$\sigma$ = 15 & RMSE = 9.24 & RMSE = 10.98\\
	\end{tabular}
	\caption{Cameraman denoising results for AWGN}
	\label{fig:cameraman_g}
\end{figure}

\begin{figure}[h]
\centering
	\begin{tabular}{ccc}
	 \textbf{Noisy} & \textbf{TV} &\textbf{QTV}\\ \hline \\
	\includegraphics[width=.245\linewidth]{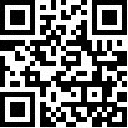} &
	\includegraphics[width=.245\linewidth]{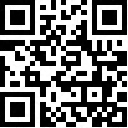} &
	\includegraphics[width=.245\linewidth]{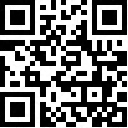} \\
	
	$\sigma$ = 5 & RMSE = 2.35 & RMSE = 2.38\\ 	%\includegraphics[width=.245\linewidth]{images/base} &
	\includegraphics[width=.245\linewidth]{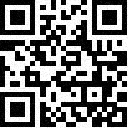} &
	\includegraphics[width=.245\linewidth]{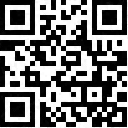} &
	\includegraphics[width=.245\linewidth]{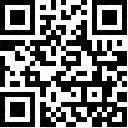} \\
	$\sigma$ = 10 & RMSE = 4.71 & RMSE = 4.71\\
	\includegraphics[width=.245\linewidth]{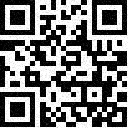} &
	\includegraphics[width=.245\linewidth]{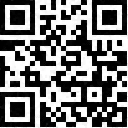} &
	\includegraphics[width=.245\linewidth]{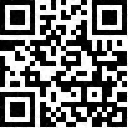} \\
	$\sigma$ = 15 & RMSE = 7.21 & RMSE = 7.22\\
	\end{tabular}
	\caption{QRCode denoising results for AWGN}
	\label{fig:qrcode_g}
\end{figure}

We consider the problem of denoising the three test images {\em lena}, {\em QR},
{\em cameraman}, corrupted only by AWG noise with standard
deviation $\sigma=\{5,10,15\}$, as shown in the first column of Fig. \ref{fig:lena_g},  Fig.\ref{fig:cameraman_g}, and Fig.\ref{fig:qrcode_g}.

The denoised images obtained by applying TV and QTV algorithms are illustrated in Fig. \ref{fig:lena_g},  Fig.\ref{fig:cameraman_g}, and Fig.\ref{fig:qrcode_g}, column-wise for the three images. For each denoised image the RMSE obtained value is reported below.

From a visual inspection the denoised images from TV and QTV present a comparable quality, even if the RMSE highlights the lost in accuracy due to the mentioned  integer arithmetic representation followed by QTV.
%lena

%cameraman

%qrcode

Moreover, the quantum TV algorithm is able to denoise images corrupted by severe Gaussian noise, as illustrated in the last row of Fig. \ref{fig:lena_g},  Fig. \ref{fig:cameraman_g}, and Fig. \ref{fig:qrcode_g} for $\sigma=15$.
Unlike, the quantum median filter proposed in \cite{QM},
as any standard median filtering, is not appropriate to remove this kind of noise. 
The median filter is instead well-known to be an excellent image denoiser in case of salt-and-pepper noise because it does not blur the image, as a mean filter would do. However, despite its name, the median filter is not a filter because it does not respect the linearity property.

\subsection{Example 2: SPN denoising}
In this example we applied TV and QTV for the  denoising of the three test images {\em lena}, {\em QR},
{\em cameraman}, corrupted by SPN noise with error rate ER$_\%=\{5,10,30\}$.

\begin{figure}[h]
\centering
	\begin{tabular}{ccc}
 \textbf{Noisy} & \textbf{TV} &\textbf{QTV}\\ \hline \\
	\includegraphics[width=.245\linewidth]{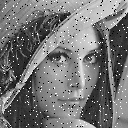} &
	\includegraphics[width=.245\linewidth]{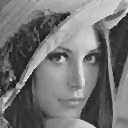} &
	\includegraphics[width=.245\linewidth]{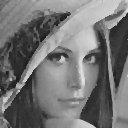} \\
	ER = 5\% & RMSE = 11.72 & RMSE = 11.95\\ 
	\includegraphics[width=.245\linewidth]{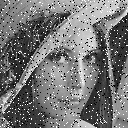} &
	\includegraphics[width=.245\linewidth]{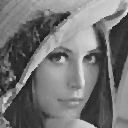} &
	\includegraphics[width=.245\linewidth]{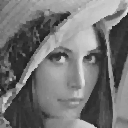} \\
	ER = 10\% & RMSE = 13.05 & RMSE = 13.28\\ %\includegraphics[width=.245\linewidth]{images/lbase} &
	\includegraphics[width=.245\linewidth]{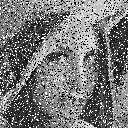} &
	\includegraphics[width=.245\linewidth]{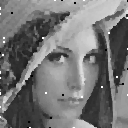} &
	\includegraphics[width=.245\linewidth]{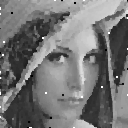} \\
	ER = 30\% & RMSE = 20.22 & RMSE = 20.71\\
	\end{tabular}
	\caption{Lena denoising results for SPN}
	\label{fig:lena_sp}
\end{figure}

\begin{figure}[h]
\centering
	\begin{tabular}{ccc}
	\textbf{Noisy} & \textbf{TV} &\textbf{QTV}\\ \hline \\
	\includegraphics[width=.245\linewidth]{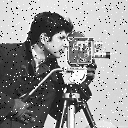} &
	\includegraphics[width=.245\linewidth]{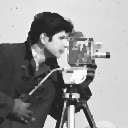} &
	\includegraphics[width=.245\linewidth]{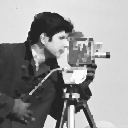} \\
	ER = 5\% & RMSE = 15.67 & RMSE = 16.05\\ 
	\includegraphics[width=.245\linewidth]{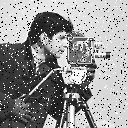} &
	\includegraphics[width=.245\linewidth]{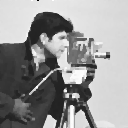} &
	\includegraphics[width=.245\linewidth]{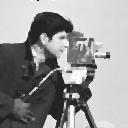} \\
	ER = 10\% & RMSE = 16.91 & RMSE = 17.56\\ %\includegraphics[width=.245\linewidth]{images/cbase} &
	\includegraphics[width=.245\linewidth]{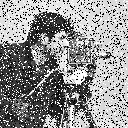} &
	\includegraphics[width=.245\linewidth]{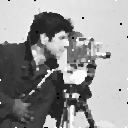} &
	\includegraphics[width=.245\linewidth]{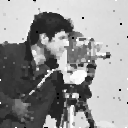} \\
	ER = 30\% & RMSE = 25.23 & RMSE = 25.75\\
	\end{tabular}
	\caption{Cameraman denoising results for SPN}
	\label{fig:cameraman_sp}
\end{figure}

\begin{figure}[h]
\centering
	\begin{tabular}{ccc}
\textbf{Noisy} & \textbf{TV} &\textbf{QTV}\\ \hline \\
	\includegraphics[width=.245\linewidth]{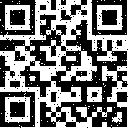} &
	\includegraphics[width=.245\linewidth]{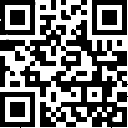} &
	\includegraphics[width=.245\linewidth]{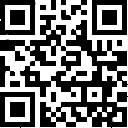} \\
	ER = 5\% & RMSE = 11.31 & RMSE = 11.46\\ 
	\includegraphics[width=.245\linewidth]{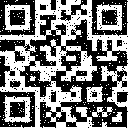} &
	\includegraphics[width=.245\linewidth]{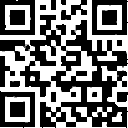} &
	\includegraphics[width=.245\linewidth]{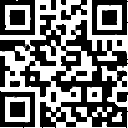} \\
	ER = 10\% & RMSE = 19.47 & RMSE = 19.50\\
	\includegraphics[width=.245\linewidth]{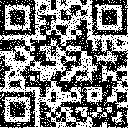} &
	\includegraphics[width=.245\linewidth]{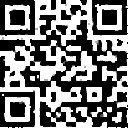} &
	\includegraphics[width=.245\linewidth]{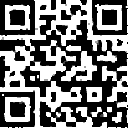} \\
	ER = 30\% & RMSE = 40.16 & RMSE = 40.25\\
	\end{tabular}
	\caption{QRCode denoising results for SPN}
	\label{fig:qrcode_sp}
\end{figure}

The noisy images are shown in the first column of
Fig.\ref{fig:lena_sp}, Fig. \ref{fig:cameraman_sp}, and Fig. \ref{fig:qrcode_sp}, together with the
denoised images in the second (TV) and third (QTV) columns, along with the associated RMSE values, reported in the bottom.

From these results, we can see how well quantum algorithm performs when compared to its variational counterpart. QTV results present excellent qualitative performance, with minimal RMSE differences. 

However, by a visual inspection of the denoised images, we notice that, for both the algorithms, some pixel clusters were not completely denoised. This  can be due to either to the limited pixel neighborhood considered (4 pixels), or to the $L_2$-norm metric fidelity in our model \eqref{eq:TVLq}, when it is well known that SPN can be better treated by a $L_1$-norm fidelity, which, anyway, leads to a non-differentiable fidelity term. 
%However this does not affect the denoising results in an appreciable way.

\section{Conclusions and discussion}
\label{sec:sec7}

In this work a quantum approach is proposed for total variation denoising, and its corresponding quantum circuit is designed. Specifically the quantum TV implements the anisotropic median formula presented in \cite{OsherTV}.
The main idea of the approach is that first
the classical image is converted into a quantum version based on the quantum
representation (NEQR) of digital images, and then three quantum modules are applied to realize the neighboring collection for each pixel in the image, weight calculation, and median extraction.
Finally, an image measurement process collapses the quantum state into a resulting denoised image.
From the complexity analysis in Section \ref{sec:sec5} we derived a polynomial complexity in the number of pixels $N$ for the NP module, and a polynomial complexity in logarithm of the number of colors $N_q=2^q$ for the PC and MF modules.

The experimental results show that the  quantum TV performance is comparable to the classical variational TV approach.
%and it is a more efficient alternative to other filters employed in Quantum Image Processing, such as the median filter described in \cite{QM}.
However, we highlighted several issues that need to be addressed to make the proposal a competitive QIP algorithm, as its variational counterpart. For example the Neighborhood collection module is an expensive operand due to the NEQR image representation.
Even though NEQR is still one of the most used representation methods in QIP and the most suitable choice for this work, future developments should definitely search for other QIR alternatives, or develop more efficient versions of the same representation, such as parametric quantum circuits that take advantage of data structure for improving image processing activities.

\bmhead{Acknowledgments}

This work is the result of a collaboration activity with CINECA High Performance Computing center, located in Bologna, Italy. This work was supported in part by the National Group for Scientific Computation (GNCS-INDAM),
Research Projects 2020, and in part by MIUR RFO projects.

\bmhead{Conflict Of Interest}

The authors declare that there are no conflicts of interest regarding the publication of this paper.

%CINECA offers support to scientific research, public and industrial, through high performance computing and the use of the most innovative supercomputing systems based on state-of-the-art architectures and technologies.

\begin{appendices}

\section{Quantum Gates for QTV}\label{secA1}
In this appendix we illustrate a few modules composed of basic quantum gates used in the design of QTV algorithm. 

\begin{figure}[b!]
\centering
\begin{tabular}{cc}
\includegraphics[width=4cm]{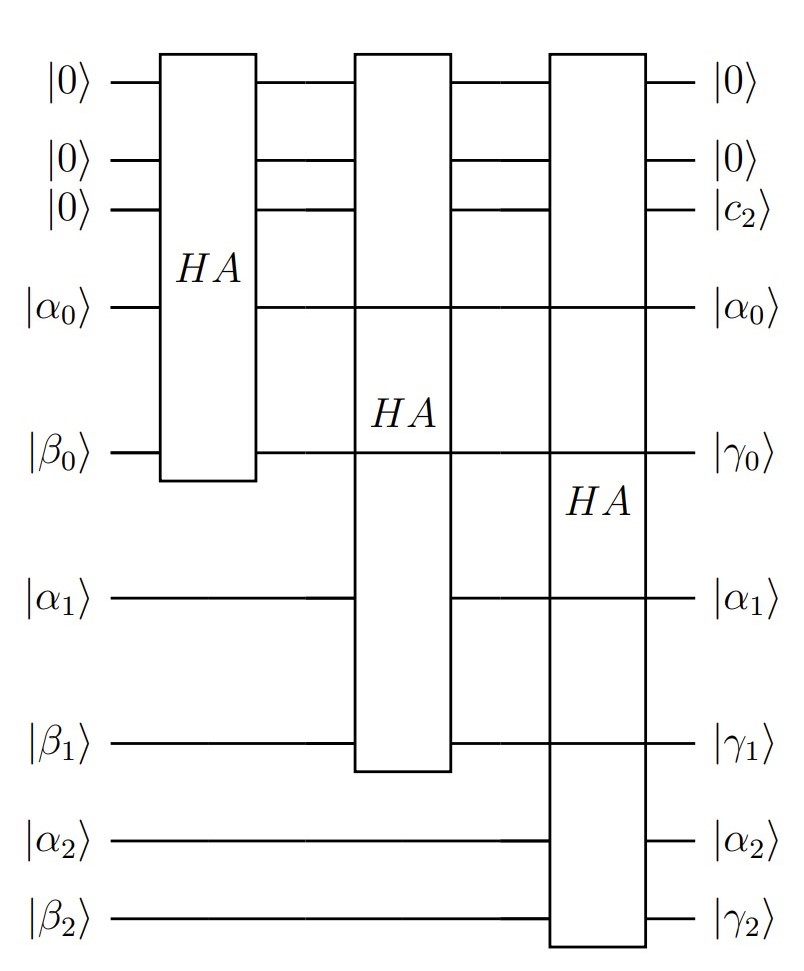} &
\includegraphics[width=7cm]{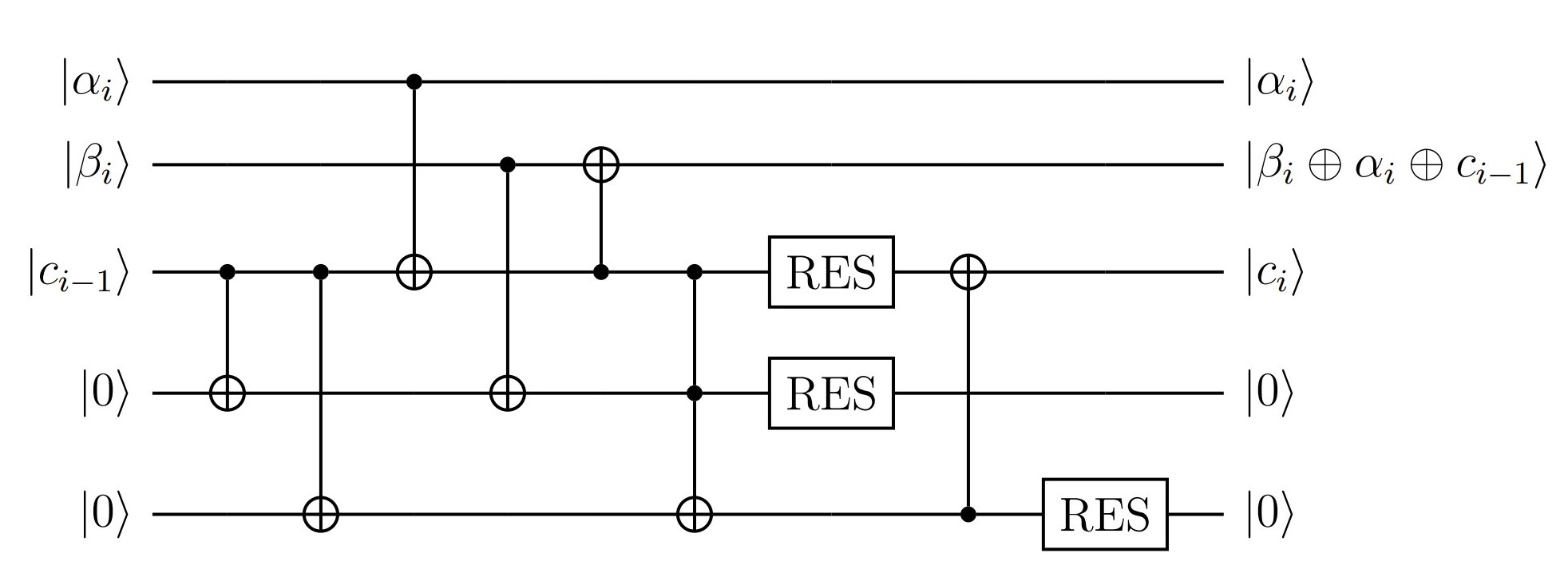}\\
(a) Full adder (FA) & (b) Half adder (HA)\\
\end{tabular}\\
\includegraphics[width=9cm]{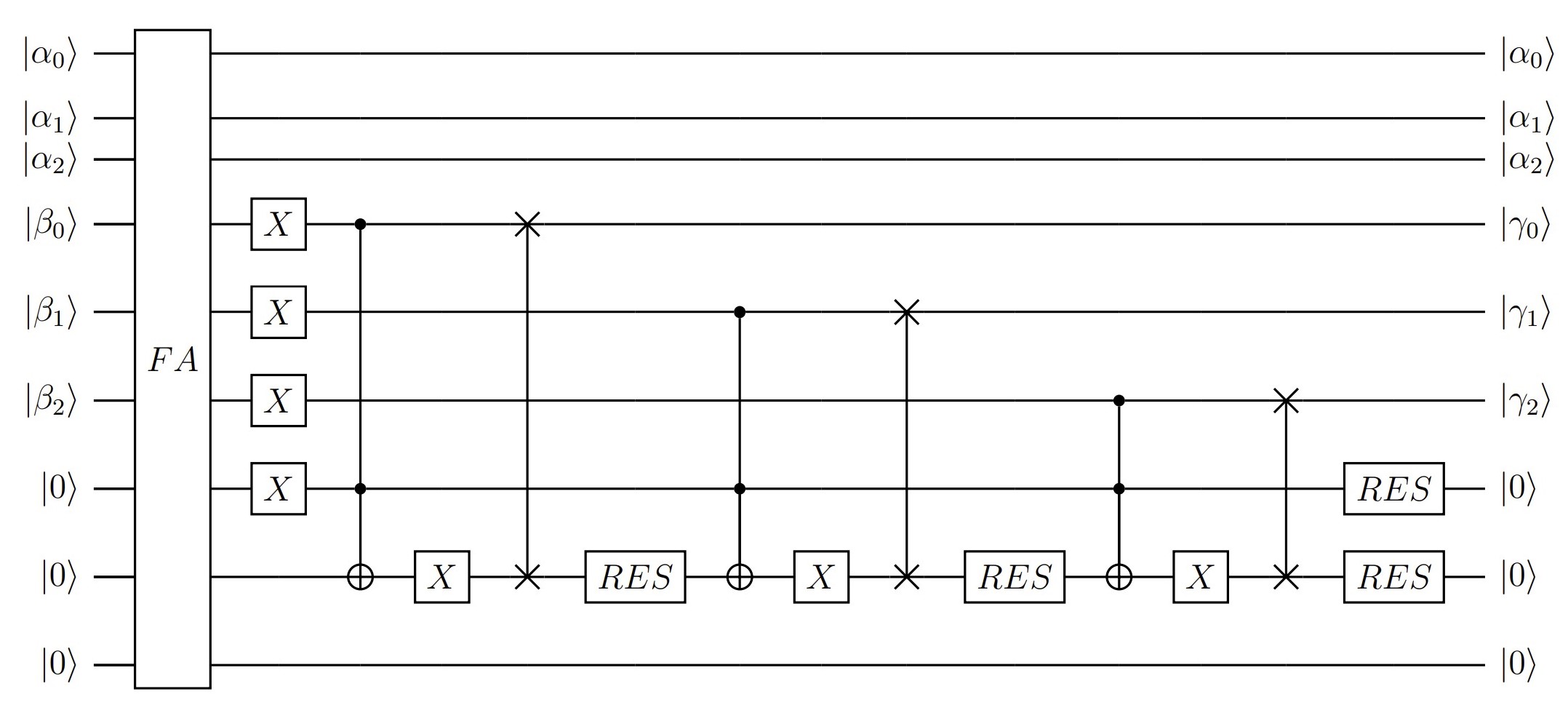}\\
(c) Adder for 3-qubits registers\\
\includegraphics[width=9cm]{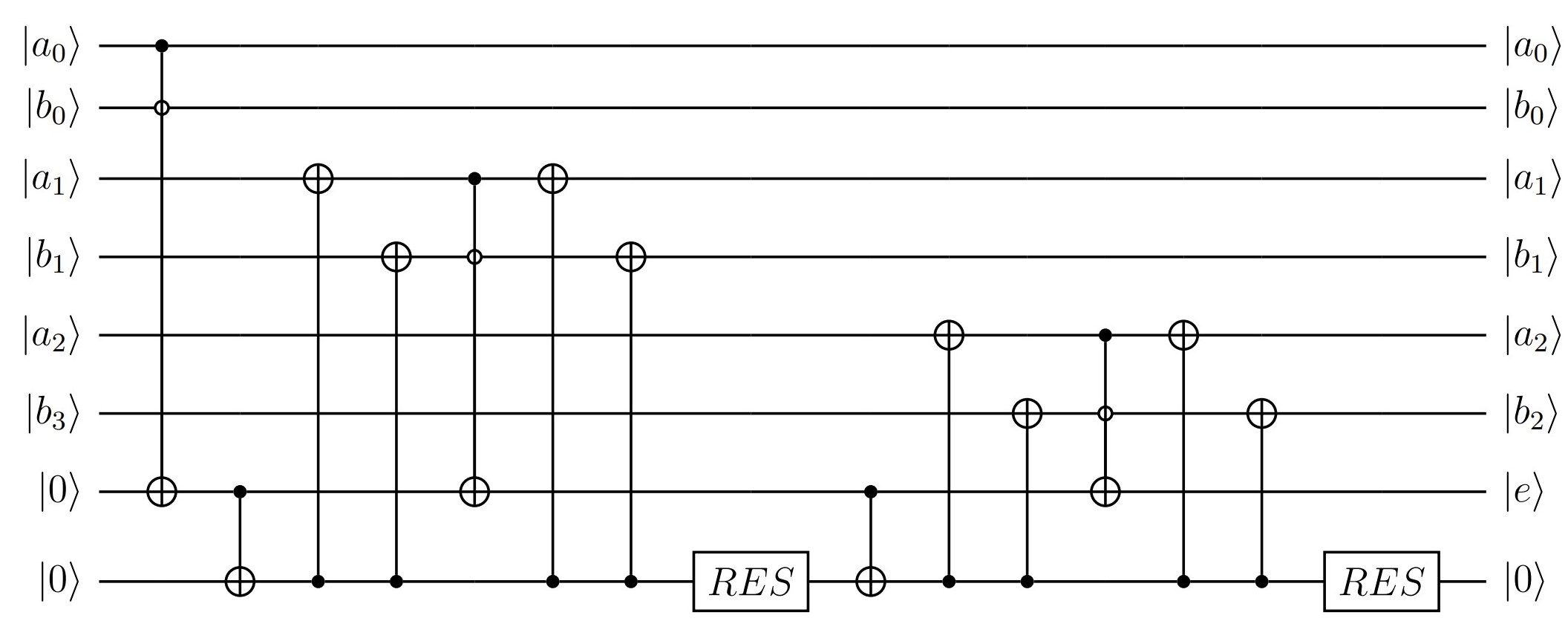},\\
(d) Comparator module  for 3-qubits registers \\
\includegraphics[width=9.0cm]{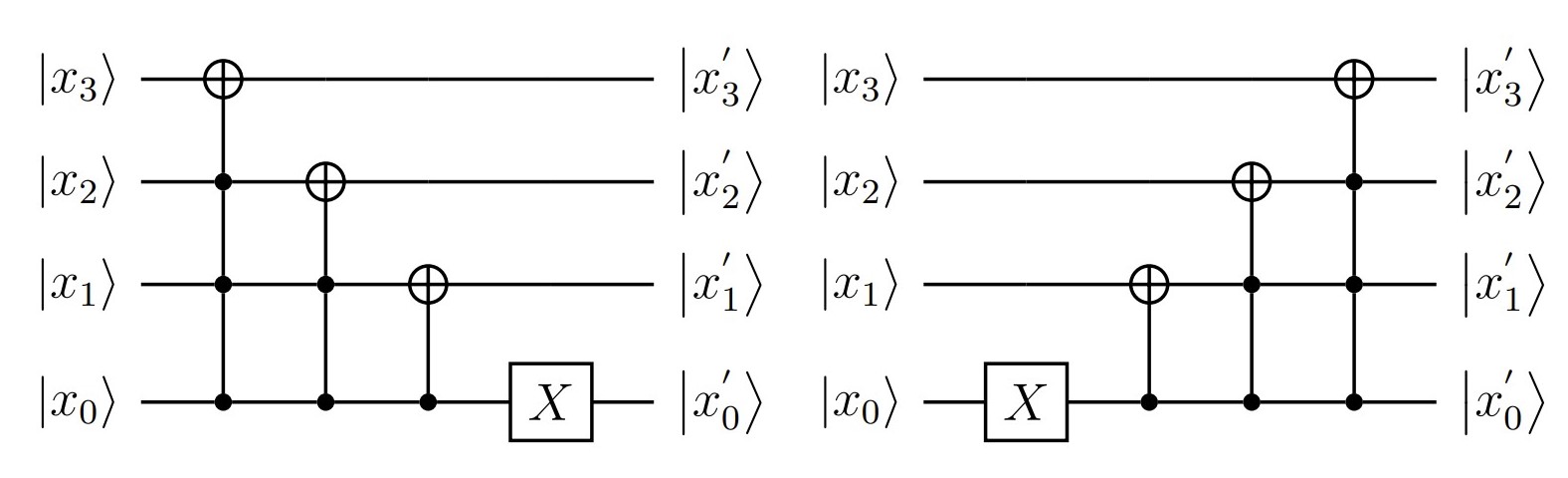}\\
(e) Cycle Shift modules: CS+ (left); CS- (right)
\caption{Quantum modules used in QTV.}
\label{fig:CM}
\end{figure}

We used classical logic for designing a quantum version of a Half-Adder (HA) and a Full-Adder (FA), as illustrated in Fig. \ref{fig:CM}(b) and \ref{fig:CM}(a) . These sub-circuits need three additional auxiliary qubits, for storing temporary results needed for summing. FA does not provide
a minimum/maximum cap over result: to solve this, we can use the temporary result stored in auxiliary qubits to manually fix the output to the correct value, as illustrated in Fig.\ref{fig:CM}(c).

The Comparator module is illustrated in Fig.\ref{fig:CM}(d). 
This module uses a very small amount of elementary gates and a reduced number of auxiliary qubits: if Reset option (RES) is not available, only $q$ ancillas are needed for Comparator to work.

A Cycle Shift module (CS) is essentially a modulo-2 adder or subtractor, defined as CS+ and CS- respectively. CS module implementation is illustrated in Fig. \ref{fig:CM}(e).

%For  a better understanding we introduce the most used formalism in quantum computing. Dirac notation (or \emph{bra-ket} notation) is a formal way to write elements and operations in a complex vector space  describing an $n$-qubit quantum system:
%\begin{itemize}
%\item 
%$\ket{0}= \left[ \begin{array}{c}
%1 \\
%0  
%\end{array} \right], \quad$
%$\ket{1}= \left[ \begin{array}{c}
%0 \\
%1  
%\end{array} \right]$
%\item
%$ \ket{\psi}  := (\psi_0, \psi_1 , ... , \psi_{N-1})^T \Rightarrow$ \emph{ket} stands for a column vector;
%\item
%$\langle \psi \rvert := (\psi_0^*, \psi_1^*, ..., \psi_{N-1}^*) \Rightarrow$ \emph{bra} stands for the conjugate transpose of the {\em ket};
%\item
%$\langle \phi \rvert \psi\rangle := \sum_{i=0}^{2^n-1} \phi_i^* \cdot \psi_i \Rightarrow$ inner product of two vectors;
%\item
%$ \ket{\phi} \otimes \ket{\psi}  \Rightarrow$ tensor product of two vectors, which results in a matrix of elements $(\phi_{i} \psi_{j})_{ij}$;
%\item $\ket{\psi}^{\otimes k} \Rightarrow$ $\ket{\psi}$ has done $k$ tensor product on itself; thus
%$\ket{0}^{\otimes k}$ is a vector of $2^{k}-1$ zeros;
%\end{itemize}

\FloatBarrier

%An appendix contains supplementary information that is not an essential part of the text itself but which may be helpful in providing a more comprehensive understanding of the research problem or it is information that is too cumbersome to be included in the body of the paper.




\end{appendices}



\end{document}